\documentclass[journal]{IEEEtran}
\usepackage{amsfonts}
\IEEEoverridecommandlockouts

\ifCLASSINFOpdf
\else
\fi
\usepackage{multirow}
\usepackage{cite}
\usepackage{stfloats}
\usepackage{color}
\usepackage{epsfig}
\usepackage{graphicx}
\usepackage{algorithm}
\usepackage{algorithmic}
\usepackage{epsf}
\usepackage{slashbox}
\usepackage{float}
\usepackage{amssymb }
\usepackage[cmex10]{amsmath}
\usepackage{stfloats}
\begin{document}
\title{
 Unilateral Left-Tail Anderson-Darling Test Based Spectrum Sensing with Laplacian Noise}
\author{\IEEEauthorblockN{Lijuan Jiang, Yongzhao Li,~\IEEEmembership{Senior Member,~IEEE},  Yinghui Ye, Yunfei Chen, Ming Jin,  Hailin Zhang,~\IEEEmembership{Member,~IEEE}}
\thanks{
Lijuan Jiang, Yongzhao Li,  Yinghui Ye  and Hailin Zhang are with the Integrated Service Networks Lab of Xidian University, Xi'an, China (e-mail:  yzli@mail.xidian.edu.cn, lijuanann@163.com, connectyyh@126.com and hlzhang@xidian.edu.cn). The corresponding author is Yinghui  Ye.
}
\thanks{Yunfei Chen is with the School of Engineering, University of Warwick, Coventry,  CV4 7AL, U.K}
\thanks{Ming Jin is with the Faculty of Electrical Engineering and Computer Science,
Ningbo University, Ningbo 315211, China. (e-mail: jinming@nbu.
edu.cn).}
\thanks{The research reported in this article was supported by  the National Key Research and Development Program of China (2016YFB1200202),    the  Natural Science Foundation of Shaanxi Province (2017JZ022), the Natural Science Foundation of China (61771365, 61501371), the 111
Project of China (B08038) and the Research Program of Education Bureau of Shaanxi Province (17JK0699).}}
\maketitle

\begin{abstract}
This paper focuses on spectrum sensing under Laplacian noise. To remit the negative effects caused by heavy-tailed behavior of Laplacian noise, the fractional lower order moments (FLOM) technology is employed to pre-process the received samples before spectrum sensing. Via exploiting the asymmetrical difference between the distribution for the FLOM of  received samples in the absence and presence of primary users, we formulate the spectrum sensing problem under Laplacian noise as a unilateral goodness-of-fit (GoF) test problem. Based on this test problem, we propose a new GoF-based detector which is called unilateral left-tail Anderson Darling (ULAD) detector. The analytical expressions for the theoretical performance, in terms of false-alarm and detection probabilities, of the ULAD are derived. Moreover, a closed-form expression for the optimal detection threshold is also derived to minimize the total error rate. Simulation results are provided to validate the theoretical analyses and to demonstrate the superior performance of the proposed detector than others.
\end{abstract}
\IEEEpeerreviewmaketitle
\section{INTRODUCTION}
\IEEEPARstart{T}{he} spectrum resource becomes increasingly scare due to the rapid growth of wireless communications. Meanwhile, report of the Federal Communication Commission  shows that a major portion of the allocated radio frequency spectrum is severely underutilized by primary users (PUs) [1]. Motivated by this, cognitive radio (CR) is proposed to alleviate this issue. In the CR domain, spectrum sensing is one of the most significant techniques, in which secondary users (SUs) can monitor the presence of PUs continuously and find available ``spectral holes'' [2].

To this end, plenty of spectrum sensing detectors, including eigenvalue-based detector, cyclostationary detector, energy detector (ED) and its modified version have been proposed [3]-[6]. These detectors are mainly derived for additive white Gaussian noise (AWGN). However, in practical communication systems, Gaussian noise usually degrades to non-Gaussian noise  due to, e.g., artificial impulsive noise, devices with electromechanical switches (printers, copy machines, electric motors in elevators, etc.), and co-channel interference from the secondary users (SUs) [7], [8]. As demonstrated in [7], [9], [10], one frequently encountered non-Gaussian noise is impulsive noise, whose distribution has an associated ``heavy-tailed'' behavior. Owing to the capability of characterizing the heavy-tailed behavior, the Laplacian noise is popular for modeling impulsive noise [7], [9]-[16]. For instance, compared to the Gaussian approximate, the Laplacian approximate is more accurate for the distribution of the multiple access interference in time-hopping ultrawide bandwidth communications [9]. The detection efficiency in [3]-[6] may degrade considerably under Laplacian noise.

To tackle the Laplacian noise, several spectrum sensing detectors have been proposed [11]-[18]. In [11], the received samples are pre-processed by a suprathreshold stochastic resonance (SSR) system before spectrum sensing. However, the SSR system involves high cost at the receivers of SUs for the summing array of a series of threshold devices. By exploiting spacial correlation among
multiple antennas, a polarity-coincidence-array (PCA) based detector was proposed [17]. Nevertheless, the PCA is applicable only in the case of multi-antenna scenario. To overcome this drawback, in [18], a soft-limited PCA (SL-PCA) based detector was proposed via replacing the sign function in PCA with a soft limiting function. With its optimum soft parameter, the SL-PCA detector can operate well in the case of a single receiver antenna.
The fractional lower order moments (FLOM) are distinguished as a powerful means in remitting the adverse effects caused by heavy-tailed  Laplacian noise [12]-[16], [19], [20]. In [12], kerenlized energy detector (KED) was proposed, which exhibits a moderate complexity. Similarly, the FLOM technology is also adopted in $p$-th order moments (POM) based detectors where $p<2$ [13], [14]. Although, with the optimum $p$ or optimum soft parameter, the detectors in [13], [14] and [18] can achieve superior performance than others, the closed-form expressions for optimum parameters have not be given. Therefore, it is necessary to adjust the parameter until find its optimum value once the sensing environment changes. However, since it is quit difficult to obtain the optimum parameters within limited sensing time, the parametric detectors in  [13], [14] and [18] may not operate well in the real-time CR system. To avoid the searching operation and reduce the computational complexity,  [15] and [16] proposed the absolute value cumulating (AVC) based detector, which is a special case of the POM-based detector by taking $p=1$.

On the other hand, the spectrum sensing problem can be formulated as a goodness-of-fit (GoF) test which can take full advantage of the statistical features of the received samples [21], [22]. Its basic principle is to test whether the empirical distribution deviates from the theoretical (expected) distribution, and then reject the null hypothesis which indicates the absence of PUs, at a certain significance level. The empirical distribution and theoretical distribution are determined by the received samples and the null hypothesis, respectively. Since the GoF test can make full use of statistical features of the received samples, it will bring a satisfactory Type \uppercase\expandafter{\romannumeral1} error and Type \uppercase\expandafter{\romannumeral2} error by selecting an appropriate fitting criterion to calculate the distance between the empirical distribution and the theoretical distribution.

Based on the GoF theory, several fitting criteria have been applied in spectrum sensing, such as the Cramervon Mises (CM) test, the Kolmogorov-Smirnov (KS) test, the Order statistic (OS) test, the Anderson Darling (AD) test, etc [21]-[31]. Nevertheless, these proposed detectors in [21]-[29] have assumed Gaussian noise. Thus it may be not appropriate to employ them directly in the presence of Laplacian noise. Although detectors in [30] and [31] are discussed in Laplacian noise, the detection performance is not satisfied. Considering the efficiency of the FLOM method, it can be combined with the GoF test to ameliorate the performance degradation caused by heavy-tailed Laplacian noise.

Motivated by above discussion, in this paper, we employ the FLOM technology to pre-process the received samples before spectrum sensing,and then formulate the spectrum sensing problem as a unilateral GoF test problem. Based on it, we propose a novel GoF test, named as unilateral left-tail AD (ULAD) test, and employ it in spectrum sensing to derive a powerful ULAD detector with low complexity. There are \emph{three reasons} for proposing the ULAD test instead of using existing GoF tests in [21]-[31] under Laplacian noise. {\color{black}Firstly, for any given $p>0$, the $p$-th order moments of received samples is monotonic increasing with samples' absolute value. Meanwhile, the power of received samples containing both PUs signals and noise (hypothesis one) is larger than that only including noise (null hypothesis) on average [25]. Hence, it can be noted that the absolute value of received samples under hypothesis one is larger than that under null hypothesis, which results in the FLOM ($0<p<2$) of received samples under hypothesis one larger than that under null hypothesis.} As a consequence, the spectrum sensing problem can be transformed into a unilateral GoF test problem, whereas those existing GoF tests used a bilateral GoF test problem. Secondly, the difference in the left tail between the theoretical distribution and the empirical distribution in the unilateral GoF test problem is larger than that in the right tail. In this case, it would be more reasonable to adopt an asymmetrical weight function (The weight function is used to evaluate the difference between the theoretical distribution and the empirical distribution. Thus, it makes a great effect on the detection performance of fitting criteria in GoF test.), which assigns a larger weight to the left tail of the theoretical distribution [32], instead of the symmetrical weight function adopted in existing GoF tests. Thirdly, ranking operations may be required in those existing GoF tests to calculate the test statistics, bringing an extra computational complexity. The main contributions of this paper are summarized as follows:
\begin{itemize}
  \item  To mitigate the heavy-tailed behavior of Laplacian noise, the FLOM technology is adopted to pre-process the received samples. By exploiting the asymmetrical difference between the theoretical distribution and empirical distribution for the FLOM of received samples, we formulate spectrum sensing problem in the presence of Laplacian noise as a unilateral GoF test problem, and then propose a low complexity and powerful ULAD detector.
  \item The analytical expressions for false alarm and detection probabilities are derived. To obtain further insights, an optimization problem in terms of the detection threshold is formulated to minimize the total error rate of the proposed detector, and its closed-form expressions is obtained.

\end{itemize}

The remainder of this paper is organized as follows. We establish the system model and formulate the spectrum sensing under Laplacian noise as a unilateral GoF test problem in Section 2. Section 3 discusses several GoF tests and proposes the low complexity and powerful ULAD detector. The performance analyses and discussions of the optimal detection threshold are given in Section 4. Section 5 shows simulation results. Finally, Section 6 concludes the paper.


\subsection{System model}\label{subsec2.1}
Let ${Y = \left\{ {{Y_i}} \right\}_{i = 1}^n}$ denote $n$ samples at instant time $i$ ($i=1,2,...,n$). For the sake of simplicity, we assume that ${{Y_i}}$ is real-valued. Thus, ${{Y_i}}$ is given as [21], [29]
\begin{equation}
\begin{array}{*{20}{l}}
{{H_0}:{Y_i} = {W_i}}\\
{{H_1}:{Y_i} = \sqrt \rho  {S_i} + {W_i}}
\end{array}
\end{equation}
where ${\rho}$ represents the signal-to-noise ratio (SNR); the primary signal $S_i$ is assumed to be binary phase-shift keyed (BPSK) with ${\sigma ^2}_s = 1$ for the sake of mathematical tractability; it also can be other signals, such as random uncorrelated Gaussian signals or the signals in sine waveform with single carrier frequency  [33]. ${{W_i}}$ is the heavy-tailed Laplacian noise with mean zero and variance ${\sigma _w^2}$, and its corresponding probability density function (PDF) is given by
\begin{equation}
{f_w}\left( y \right) = \frac{1}{{\sqrt {2\sigma _w^2} }}{\rm{exp}}\left[ { - \sqrt {\frac{2}{{\sigma _w^2}}} \left| y \right|} \right].
\end{equation}
Without loss of generality, we assume that the
noise samples are independent, identically distributed (i.i.d.),
and are independent of the primary signal ${S_i}$. Since the samples depend on noise under $H_0$ or both the noise and primary signals under $H_1$, ${Y_1},{Y_2}, \cdots ,{Y_n}$ can be regarded as i.i.d. sequences.

\newcounter{mytempeqncnt}
\begin{figure*}[t]
\normalsize
\setcounter{mytempeqncnt}{\value{equation}}
\setcounter{equation}{3}
\begin{align}
{f_1}\left( x \right) = f\left( {\left| {\sqrt \rho   + {W_i}} \right|} \right) = f\left( {\left| {{W_i} - \sqrt \rho  } \right|} \right) = \left\{ {\begin{array}{*{20}{c}}
{\frac{1}{{\sqrt {2\sigma _w^2} }}\exp \left( { - \sqrt {\frac{2}{{\sigma _w^2}}} \left( {\sqrt \rho   - x} \right)} \right) + \frac{1}{{\sqrt {2\sigma _w^2} }}\exp \left( { - \sqrt {\frac{2}{{\sigma _w^2}}} \left( {x + \sqrt \rho  } \right)} \right),0 \le x \le \sqrt \rho  }\\
{\frac{1}{{\sqrt {2\sigma _w^2} }}\exp \left( { - \sqrt {\frac{2}{{\sigma _w^2}}} \left( {x - \sqrt \rho  } \right)} \right) + \frac{1}{{\sqrt {2\sigma _w^2} }}\exp \left( { - \sqrt {\frac{2}{{\sigma _w^2}}} \left( {x + \sqrt \rho  } \right)} \right),\;\;\;\;\;\;x > \sqrt \rho  }
\end{array}} \right.
\end{align}
\setcounter{equation}{4}
\hrulefill
\end{figure*}

\subsection{GoF test problem}\label{sec2.2}
By selecting an exponent varying from 0 to 2, the FLOM is powerful in mitigating the negative effects caused by heavy-tailed behavior of  Laplacian noise [12]-[16], [19], [20]. Besides, in terms of FLOM technology, the computational complexity of non-integer exponents is larger than that of integer exponents [14]. Hence, in order to reduce computational complexity, in this subsection, we adopt the absolute value of $Y_i$,  a
special form of FLOM with the lowest complexity, as our observation, i.e., ${{x_i} = \left| {{Y_i}} \right|}$, ($i=1,2,...,n$).

If the null hypothesis ${{H_0}}$ is accepted, the sample ${Y_i}$ is drawn with the noise (Laplacian) distribution and the observation ${{x_i}}$ follows the exponential distribution with parameter ${\sqrt {\frac{{\sigma _w^2}}{2}} }$ [15]. The corresponding PDF ${{f_0}\left( x \right)}$ can be written as
\begin{equation}
 \setcounter{equation}{3}
{f_0}\left( x \right) = \left\{ {\begin{array}{*{20}{c}}
{\sqrt {\frac{2}{{\sigma _w^2}}} \exp \left( { - \sqrt {\frac{2}{{\sigma _w^2}}} x} \right),\;\;x \ge 0}\\
{0\;\;\;\;\;\;\;\;\;\;\;\;\;\;\;\;\;\;\;\;\;\;\;\;\;\;\;\;\;\;\;,{\rm{others}}}.
\end{array}} \right.
\end{equation}
On the contrary, if the null hypothesis ${{H_0}}$ is rejected, the distribution of ${{x_i}}$ deviates from the exponential distribution ${f_0}\left( x \right)$. The PDF of ${{x_i}}$ can be expressed as (4), shown at the top of the page. Notice that the equation (4) is valid due to $\Pr \left( {{S_i} =  + 1} \right) = \Pr \left( {{S_i} =  - 1} \right) = \frac{1}{2}$. Thus, {\color{black}according to the Glivenko-Cantelli theorem [34]}, the spectrum sensing can be formulated as a typical GoF test problem, given by
 \begin{equation}
 \setcounter{equation}{5}
 \begin{array}{l}
{H_0}:{F_n}\left( x \right) = {F_0}\left( x \right)\\
{H_1}:{F_n}\left( x \right)  \ne  {F_0}\left( x \right)
\end{array}
\end{equation}
where the theoretical cumulative distribution function ${{F_0}\left( x \right)}$ is obtained by integrating $f_0(x)$ in (3), given as
\begin{equation}
{F_0}\left( x \right) = 1 - \exp \left( { - \sqrt {\frac{2}{{\sigma _w^2}}} x} \right),x \ge 0;
\end{equation}
the empirical distribution ${{F_n}\left( x \right)}$ is obtained from $x_i$ as
 \begin{equation}
{F_n}\left( x \right) = |\left\{ {i:{x_i} \le x,1 \le i \le n} \right\}|/n
\end{equation}
with ${| \bullet |}$ indicating cardinality.

By carefully examining the statistical features of $x_i$ between $H_0$ and $H_1$, there are two significant points as follows.

(i) Recognizing the fact that the value of $x_i$ in the presence of both noise and primary signals is statistically larger than that under ${H_0}$, we reformulate the above GoF test problem (${H_1}:{F_n}\left( x \right) < {F_0}\left( x \right)\;{\rm{or}}\;{F_n}\left( x \right) > {F_0}\left( x \right)$) as a unilateral GoF test problem (${H_1}:{F_n}\left( x \right) < {F_0}\left( x \right)$), that is
\begin{equation}
 \setcounter{equation}{8}
 \begin{array}{l}
{H_0}:{F_n}\left( x \right) = {F_0}\left( x \right)\\
{H_1}:{F_n}\left( x \right) < {F_0}\left( x \right).
\end{array}
\end{equation}

(ii) Fig. 1 shows the PDFs of $x_i$ under $H_0$ and $H_1$ from (3) and (4), respectively. It can be observed that ${f_1}\left( x \right)$ and ${f_0}\left( x \right)$ have larger difference at the \emph{left tail} ( the area that satisfies ${{F_0}\left( x \right) \le 0.5}$ is divided as the left tail of the theoretical distribution; otherwise belongs to the right tail). Besides, the difference increases along with the received SNR.

On the basis of above discussion, spectrum sensing problem in (1) is now equivalent to test the unilateral GoF test problem in (8) with focusing more on the difference in left tail of the theoretical distribution.

\begin{figure}[!t]
  \centering{\includegraphics[width=0.4\textwidth]{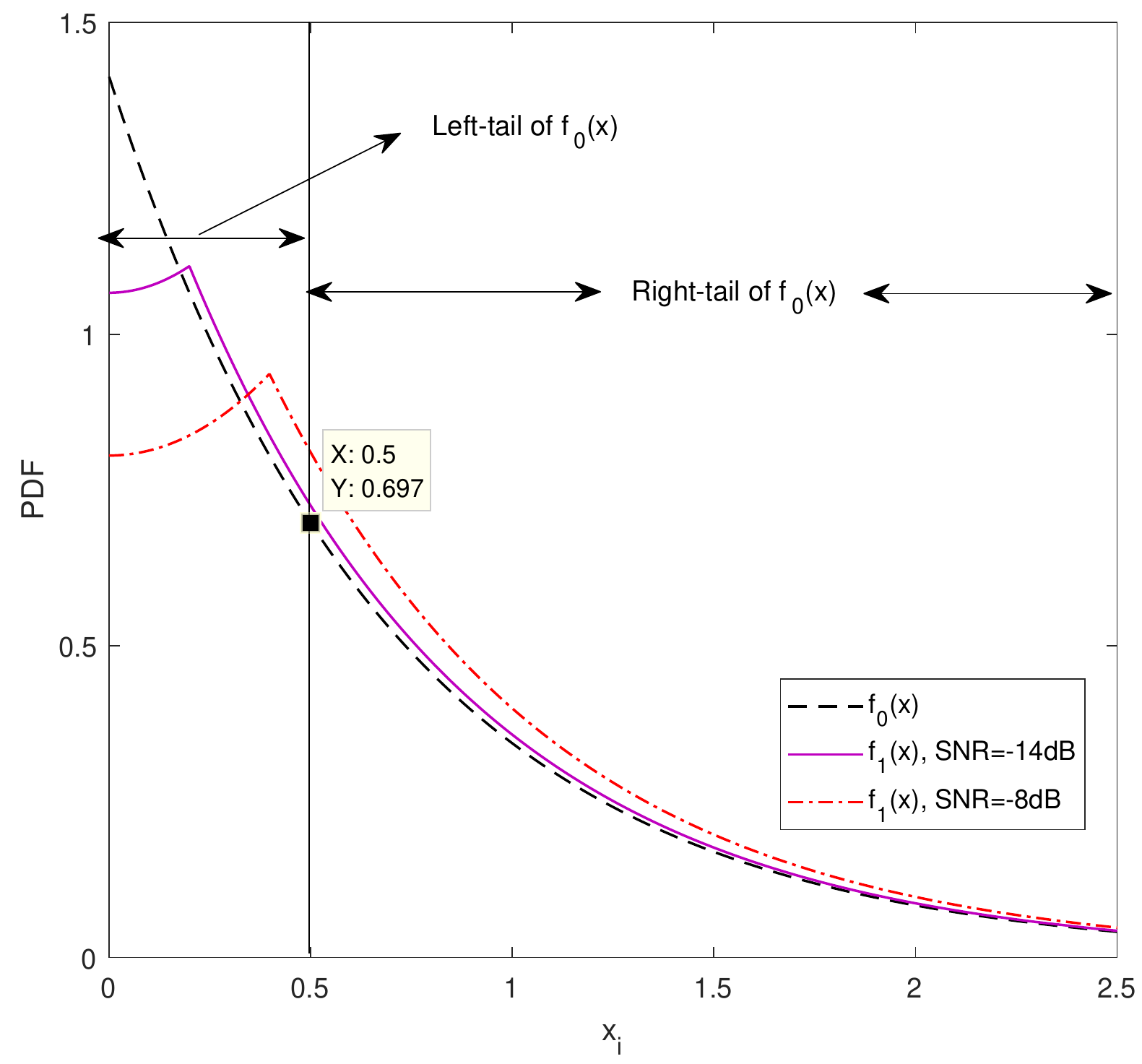}}
  \caption{Empirical and theoretical PDFs of ${{x_i}}$ with $\sigma _w^2=1$ and $n=1000$}
  \label{Fig. 1}
  \end{figure}

\section{The ULAD detector}\label{sec3}
In this section, we discuss several classical GoF tests and then propose a powerful GoF detector, named as ULAD detector.

\subsection{Previous works}\label{subsec3.1}
Several GoF tests  have been studied in literature [21]-[30], mainly including the CM test, KS test and AD test.

In KS test [30], the distance between ${F_0}\left( x \right)$ and ${F_n}\left( x \right)$ is defined as
\begin{equation}
{D_n} = \max \left| {{F_n}\left( x \right) - {F_0}\left( x \right)} \right|.
\end{equation}

In CM test [22], the received samples are arranged in an ascending order (i.e., ${x_{\left( 1 \right)}} < {x_{\left( 2 \right)}} <  \cdots  < {x_{\left( n \right)}}$), and the distance between ${F_0}\left( x \right)$ and ${F_n}\left( x \right)$ is defined as
\begin{align}\notag
W_N^2 &= n\int_{ - \infty }^{ + \infty } {{{\left[ {{F_n}\left( x \right) - {F_0}\left( x\right)} \right]}^2}} d{F_0}\left( x \right)\\
&= \sum\limits_{i = 1}^n {{{\left[ {{z_{\left( i \right)}} - \left( {2i - 1} \right)/2n} \right]}^2}}  + \left( {1/12n} \right)
\end{align}
where ${z_{\left( i \right)}} = {F_0}\left( {{x_{\left( i \right)}}} \right)$.

Different from CM test, the AD test statistic $A_N^2$ gives a symmetrical weight to both tails of ${{F_0}\left( x \right)}$, and it is expressed as [21]
\begin{align}\notag
A_N^2 &= n{\int_{ - \infty }^{ + \infty } {\left[ {{F_n}\left( x \right) - {F_0}\left( x \right)} \right]} ^2}\frac{{d{F_0}\left( x \right)}}{{{F_0}\left( x \right)\left( {1 - {F_0}\left( x \right)} \right)}}\\
&=  - \frac{{\sum\limits_{i = 1}^n {\left( {2i - 1} \right)\left( {\ln {z_{\left( i \right)}} + \ln \left( {1 - {z_{n + 1 - {\left( i \right)}}}} \right)} \right)} }}{n} - n.
\end{align}
where ${z_{\left( i \right)}} = {F_0}\left( {{x_{\left( i \right)}}} \right)$.
Even though above fitting criteria have been introduced in spectrum sensing, it is not appropriate to use them directly in our system model, and \emph{the reasons} can be concluded as follows.

{\textbf{Reason 1.}} The spectrum sensing in [21], [22] and [30] is formulated as a bilateral GoF test problem as in (5), without fully considering the statistical features that in the presence of Laplacian noise, the observations $x_i=\left| {{Y_i}} \right|$ under $H_1$ is statistically larger than that under $H_0$. As discussed in Subsection 2.2, in this case, it would be more appropriate to formulate the spectrum sensing as a unilateral GoF test problem.

{\textbf{Reason 2.}} The weight function, used to evaluate the difference between ${{F_n}\left( x \right)}$ and ${{F_0}\left( x \right)}$, is vital to the detection performance of GoF test. The weight functions of KS and CM are the continuous uniform distribution with parameters 0 and 1, which give an equal weight to both tails of the theoretical distribution. The symmetrical weight function of AD, ${{1 \mathord{\left/{\vphantom {1 {\left\{ {{F_0}\left( x \right)\left( {1 - {F_0}\left( x \right)} \right)} \right\}}}} \right.\kern-\nulldelimiterspace} {\left\{ {{F_0}\left( x \right)\left( {1 - {F_0}\left( x \right)} \right)} \right\}}}}$, assigns a symmetrical weight to both tails of ${{F_0}\left( x \right)}$. Apparently, the uniformly distributed or symmetrical weight functions in the KS, CM and AD are not the best candidate for the unilateral GoF test problem. The reason is that if the difference focuses more on the left tail (or right tail) of the theoretical distribution, an effective strategy for improving the detection performance is to adopt an asymmetrical weight function to assign a larger weight to the left tail (or right tail) [32].
Considering the unilateral GoF test problem, one can observe from Fig. 1 is that the difference focuses more on the left tail. Accordingly,  a properly designed asymmetrical weight function with focusing on the left tail would be able to offer superior detection  performance  instead of the existing ones in [21], [22] and [30].

{\textbf{Reason 3.}} Due to the factor ${{\left( {2i - 1} \right)}}$ related to ${z_{\left( i \right)}}$ in (10) and (11), the CM and AD involve high computational complexity ${O\left( {n{{\log }_2}n} \right)}$ of ranking the observations in prior.

To sum up, employing these classical GoF tests directly in the presence of Laplacian noise may result in performance degradation while with high complexity.

\subsection{The ULAD test and ULAD detector}\label{subsec3.2}
In this subsection, we propose a new ULAD test under Laplacian noise and employ it in spectrum sensing to proposed the powerful ULAD detector.

According to the discussion in Subsection 3.1, we adopt an asymmetrical weight function, $1/{F_0}\left( x \right)$, to assign a larger weight to the left tail of ${F_0}\left( x \right)$ and the constructed test statistic $B_n$ can be given as
\begin{equation}
B_n = n\int_{ - \infty }^{ + \infty } {\left[ {{F_0}\left( x \right) - {F_n}\left( x \right)} \right]\frac{{d{F_0}\left( x \right)}}{{{F_0}\left( x \right)}}}.
\end{equation}

{\bf{Remark 1.}} $1/{F_0}\left( x \right)$ is a classical asymmetrical weight function in mathematics to emphasize the distance ${{F_0}\left( x \right)}-{{F_n}\left( x \right)}$ in the left tail [32].

{\bf{Remark 2.}} It is appealing that we can derive a closed-form expression of $B_n$ easily with such asymmetrical weight, shown in (13).

By breaking the integral in (12) into $n$ parts, $B_n$ can be simplified as

\begin{align}\notag
{B_n} &= n\int_{ - \infty }^{ + \infty } {\left[ {{F_0}\left( x \right) - {F_n}\left( x \right)} \right]} \frac{{d{F_0}\left( x \right)}}{{{F_0}\left( x \right)}}\\\notag
 &= n\int_{ - \infty }^{{x_1}} {\frac{{{F_0}\left( x \right)}}{{{F_0}\left( x \right)}}d{F_0}\left( x \right)} \\\notag
 &+ n\int_{{x_1}}^{{x_2}} {\frac{{{F_0}\left( x \right) - 1/n}}{{{F_0}\left( x \right)}}d{F_0}\left( x \right)}\\\notag
 & +  \cdots \\\notag
 & + n\int_{{x_{n - 1}}}^{{x_n}} {\frac{{{F_0}\left( x \right) - \left( {\left( {n - 1} \right)/n} \right)}}{{{F_0}\left( x \right)}}d{F_0}\left( x \right)} \\\notag
 & + n\int_{{x_n}}^{ + \infty } {\frac{{{F_0}\left( x \right) - \left( {n/n} \right)}}{{{F_0}\left( x \right)}}d{F_0}\left( x \right)} \\\notag
 &= n + \sum\limits_{i = 1}^n {\ln {z_{\left( i \right)}}}\\
 &\mathop  = \limits^{\left( a \right)} n + \sum\limits_{i = 1}^n {\ln {z_i}}
\end{align}
where ${{z_i} = {F_0}\left( {{x_i}} \right)}$, and the step (a) is valid since the sorting operation of $x_i$ is unnecessary in ULAD test.

By means of  the ULAD test, we derive a low complexity and powerful ULAD detector as
\begin{equation}
B_n = n + \sum\limits_{i = 1}^n {\ln {z_i}} \begin{array}{*{20}{c}}
{\mathop  \ge \limits^{{H_1}} }\\
{\mathop  < \limits_{{H_0}} }
\end{array}{\gamma _{{\rm{ULAD}}}}
\end{equation}
where ${\gamma _{{\rm{ULAD}}}}$ represents the detection threshold. If $B_n< {\gamma _{{\rm{ULAD}}}}$, we accept the null hypothesis $H_0$; otherwise, $H_0$ hypothesis is rejected.

\section{Performance analyses and optimal detection threshold}\label{sec4}
In this section, we carry out the detection performance analyses and complexity analysis of the proposed ULAD detector. Also, the closed-form expression for the optimal detection threshold is derived to minimize the total error rate.

\subsection{Detection performance analyses}\label{subsubsec4.1}
In CR, the sensing performance can be evaluated with two probabilities: probability of detection ${P_d}$ and probability of false alarm ${P_f}$, given by
\begin{align}
&{{P_d}{\rm{ = Prob}}\left\{ {B_n \ge {\gamma _{{\rm{ULAD}}}}{\rm{|}}{H_1}} \right\}}  \\
&{P_f}{\rm{ = Prob}}\left\{ {B_n \ge {\gamma _{{\rm{ULAD}}}}{\rm{|}}{H_0}} \right\}.
 \end{align}
In the following, we will derive the analytical expressions for ${P_d}$ and ${P_f}$ by means of the central limit theorem.

As stated in system model, the samples ${Y_1},{Y_2}, \cdots ,{Y_n}$ are i.i.d. sequences, then ${{x_1},{x_2}, \cdots ,{x_n}}$ are i.i.d sequences because of ${{x_i} = \left| {{Y_i}} \right|}$. For ${{z_i} = {F_0}\left( {{x_i}} \right)}$, ${{z_1},{z_2}, \cdots ,{z_n}}$ can be regarded as i.i.d sequences. Therefore, ${\ln {z_1},\ln {z_2}, \cdots, \ln {z_i}}$ are also i.i.d sequences. By means of the central limit theorem, we approximate the test statistic ${{B_n}}$ (under ${{H_0}}$ and ${{H_1}}$) as the following Gaussian distribution when the number of samples $n$ is large enough.
\begin{equation}
{B_n} \sim N\left( {n + n \times E\left[ {\ln {z_i}} \right],n \times D\left[ {\ln {z_i}} \right]} \right)
\end{equation}
where ${{E\left[ {\ln {z_i}} \right]}}$ and ${{D\left[ {\ln {z_i}} \right]}}$ represent the mean and variance of the random variable ${\ln {z_i}}$, respectively.

Based on (17), we can obtain the analytical expressions of ${P_d}$ and ${P_f}$, by analyzing the mean and variance of ${\ln {z_i}}$ in the presence and absence of primary signals, respectively. Therefore, several Propositions are given in the following.

{\textbf{Proposition 1:}} When ${H_0}$ is accepted, the mean and second-order origin moment of the variable ${\ln {z_i}}$ are given by
\begin{equation}
E\left[ {\ln {z_i}|{H_0}} \right]  =  - 1
\end{equation}
\begin{equation}
E\left[ {{{\left( {\ln {z_i}} \right)}^2}|{H_0}} \right] = 2.
\end{equation}

\emph{Proof.} The proof of Proposition 1 is detailed in Appendix-A.

Based on the Proposition 1, the variance of ${\ln {z_i}}$ can be calculated by
\begin{equation}
D\left[ {\ln {z_i}|{H_0}} \right] = E\left[ {{{\left( {\ln {z_i}} \right)}^2}|{H_0}} \right] - {\left( {E\left[ {\ln {z_i}|{H_0}} \right]} \right)^2} = 1.
\end{equation}
By substituting (18) and (20) into (17), the false alarm probability of proposed ULAD detector is expressed as
\begin{equation}
{P_f} = Q\left( {\frac{{{\gamma _{{\rm{ULAD}}}}}}{{\sqrt n }}} \right)
\end{equation}
where ${Q\left( x \right) = \frac{1}{{\sqrt {2\pi } }}\int_x^{ + \infty } {{\rm{exp}}\left( { - \frac{{{t^2}}}{2}} \right)} dt}$.

Thus, for any given ${{P_f}}$, the detection threshold ${\gamma _{{\rm{ULAD}}}}$ can be given as
\begin{equation}
{\gamma _{{\rm{ULAD}}}} = {Q^{ - 1}}\left( {{P_f}} \right)  \sqrt n.
\end{equation}

{\textbf{Proposition 2:}} When ${H_0}$ is rejected,   the mean and the second-order origin moment of ${\ln {z_i}}$ are given by
\begin{align}\notag
E\left[ {\ln {z_i}|{H_1}} \right]&= \frac{q}{2}\left[ {\frac{{\ln C}}{{1 - C}} - \ln \left( {\frac{C}{{1 - C}}} \right)} \right] \\
&+ \frac{1}{{2q}}\left[ {C - C\ln C - 1} \right] - \frac{q}{2}
\end{align}
\begin{align}\notag
 E\left[ {{{\left( {\ln {z_i}} \right)}^2}|{H_1}} \right]&= \frac{{C\left( {q - 1} \right)}}{{2q}}{\left( {\ln C} \right)^2} + q\ln C\ln q \\
 &+ \frac{C}{q}\ln C + q\mathop {\lim }\limits_{k \to \infty } \sum\limits_{i = 1}^k {\frac{{{C^i}}}{{{i^2}}}}  + 1 + q
\end{align}
where
\begin{align} \notag
&q = \exp \left( { - \sqrt {\frac{{2\rho }}{{\sigma _w^2}}} } \right)\\ \notag
&C = 1 - q.
\end{align}

\emph{Proof.} The proof of Proposition 2 is detailed in Appendix-B.

Based on the Proposition 2, the accurate variance of ${\ln {z_i}}$ under ${H_1}$ is given by
\begin{equation}
D\left[ {\ln {z_i}|{H_1}} \right] = E\left[ {{{\left( {\ln {z_i}} \right)}^2}|{H_1}} \right] - {\left( {E\left[ {\ln {z_i}|{H_1}} \right]} \right)^2}.
\end{equation}
By substituting (23) and (25) into (17), for a given detection threshold ${\gamma _{{\rm{ULAD}}}}$, the analytical detection probability is expressed as
\begin{equation}
{P_d} = Q\left( {\frac{{{\gamma _{{\rm{ULAD}}}} - n - n \times E\left[ {\ln {z_i}|{H_1}} \right]}}{{\sqrt {n \times D\left[ {\ln {z_i}|{H_1}} \right]} }}} \right).
\end{equation}
Note that there is no closed-form expression of $P_d$ due to the infinite series involved in the analytical expression in (24), however, the infinite series can be calculated by simulation. Instead of finding the
closed-form expression, we derive an approximate expression for $P_d$ in what follows.

It can be obtained that ${0 < q = \exp \left( { - \sqrt {\frac{{2\rho }}{{\sigma _w^2}}} } \right)}< 1 $, then the inequality $0<(C=1-q) <1$ holds. As ${i^2} \ge {i^0} $, we have
\begin{equation}
\mathop {\lim }\limits_{k \to \infty } \sum\limits_{i = 1}^k {\frac{{{C^i}}}{{{i^2}}}}  < \mathop {\lim }\limits_{k \to \infty } \sum\limits_{i = 1}^k {{C^i}}  = \frac{C}{{1 - C}},\left( {\left| C \right| < 1} \right)
\end{equation}
where the equality is derived from the Eq.(0.231) in [35].
Hence, by replacing ${\mathop {\lim }\limits_{k \to \infty } \sum\limits_{i = 1}^k {\frac{{{C^i}}}{{{i^2}}}} }$ in (24) with its upper bound $\frac{C}{{1 - C}}$, the approximate variance $\tilde D\left[ {\ln {z_i}|{H_1}} \right]$ and the approximate detection probability ${{\tilde P}_d}$ can be written as
\begin{align}
&\tilde D\left[ {\ln {z_i}|{H_1}} \right] = \tilde E\left[ {{{\left( {\ln {z_i}} \right)}^2}|{H_1}} \right] - {\left( {E\left[ {\ln {z_i}|{H_1}} \right]} \right)^2}\\
&{{\tilde P}_d} = Q\left( {\frac{{{\gamma _{{\rm{ULAD}}}} - n - n \times E\left[ {\ln {z_i}|{H_1}} \right]}}{{\sqrt {n \times \tilde D\left[ {\ln {z_i}|{H_1}} \right]} }}} \right)
\end{align}
where
\begin{align}\notag
\tilde E\left[ {{{\left( {\ln {z_i}} \right)}^2}|{H_1}} \right] &= \frac{{C\left( {q - 1} \right)}}{{2q}}{\left( {\ln C} \right)^2} + q\ln q\ln C\\\notag
& + \frac{C}{q}\ln C + 2.
\end{align}

By substituting (22) into  (26) or (29), the ROC curves for the proposed detector can be further obtained.

Detailed procedure of the ULAD detector is summarized in Algorithm $1$. Note that the noise variance ${\sigma _w^2}$ is assumed to be known in this paper as in [4], [13]-[16]. Even though, the noise uncertainty may affect the performance of the proposed detector, it is beyond the scope of this paper.

\begin{algorithm}
\caption{ULAD Detector with Given $P_f$}
\label{alg1}
\hspace*{0.02in} {\bf Input:} 
The number of samples $n$ and noise variance ${\sigma _w^2}$ \\
\hspace*{0.02in} {\bf Output:} 
Sensing result ${{{H_0}} \mathord{\left/
 {\vphantom {{{H_0}} {{H_1}}}} \right.
 \kern-\nulldelimiterspace} {{H_1}}}$
\begin{algorithmic}[1] 
\STATE Sample the received signals on the interest band at instant time $i$ ($i=1,2,...,n$); 
\STATE Take the absolute value of the signal sample $Y_i$ as the observation, i.e.,  ${{x_i} = \left| {{Y_i}} \right|}$;
\STATE Calculate $z_i$ using (6);
\STATE Calculate the test statistic $B_n$ according to (13);
\STATE Compute the detection threshold ${\gamma _{{\rm{ULAD}}}}$ based on (22) for the given $P_f$;
\IF{$B_n \ge {\gamma _{{\rm{ULAD}}}}$}
\STATE Reject $H_0$. The primary user is present;
\ELSE
\STATE Accept $H_0$. The primary user is absent;
\ENDIF
\end{algorithmic}
\end{algorithm}

\subsection{Analysis of complexity}\label{subsec4.2}

The computational complexity of several sensing detectors, i.e., ULAD, POM, and so on, are given in Table 1. It can be seen that the proposed ULAD, AVC, POM, KS and ED have the lowest complexities than other detectors. For AD and CM, the requirement of ranking operation brings an extra complexity with ${O\left( {n{{\log }_2}n} \right)}$. On the other hand, the proposed ULAD have the lowest complexity ${O(n)}$ due to the unnecessary of ranking operation. Since the AVC, ED and POM only sum the $p$-th order moments ($p$=1, 2, $<2$) of received samples to construct their test statistics, they have the same complexity with ${O(n)}$.

\begin{table}[!t]
\centering
\caption{The complexity of sensing algorithms}
\begin{tabular}{|c|c|}
\hline
 sensing algorithm & complexity  \\
\hline
  ULAD & ${O(n)}$ \\
\hline
  AD & ${O\left( {n{{\log }_2}n} \right)}$ \\
\hline
  CM & ${O\left( {n{{\log }_2}n} \right)}$ \\
\hline
  KS & ${O(n)}$ \\
\hline
  AVC & ${O(n)}$ \\
\hline
  ED & ${O(n)}$ \\
\hline
\end{tabular}
\end{table}

\subsection{Optimal detection threshold}\label{subsec4.2}
As we know, for a powerful sensing detetctor, a large detection probability is desired for a fixed false alarm probability. According to (15) and (16), $P_d$ and $P_f$ depend on the detection threshold ${\gamma _{{\rm{ULAD}}}}$. Thus, it is important to find an optimal detection threshold to balance ${P_d}$ and ${P_f}$.

By minimizing the total error rate, an optimization problem in terms of detection threshold is formulated, which is subject to a constraint on the false alarm probability. Such optimization problem is reasonable since the detection and false alarm probabilities are both considered in total error rate. The optimization problem can be described as [36]
\begin{subequations}
\begin{align}
&\mathop {{\rm{min}}}\limits_{{\gamma _{{\rm{ULAD}}}}}\; {P_{{\rm{error}}}}\\
&{\rm{s}}{\rm{.t}}.\;\;{P_f} \le {\zeta _{{P_f}}}
\end{align}
\end{subequations}
where ${P_{\rm{error}}} = {P_f} + \left( {1 - {P_d}} \right)$ is total error rate,  ${{\zeta _{{P_f}}}}$ represents the threshold of false alarm probability. Here, we set ${\zeta _{{P_f}}}=$0.1 since $P_f$ is specified to be not larger than 0.1 to guarantee the opportunities of assess to the vacant spectrum in CR systems [37].

Based on the expressions of $P_f$ in (21) and ${{\tilde P}_d}$ in (29), the first order derivative of ${{P_{\rm{error}}}}$ is given as
\begin{equation}
\frac{{\partial {P_{{\rm{error}}}}}}{{\partial {\gamma _{{\rm{ULAD}}}}}} =  - \frac{1}{{\sqrt {2n\pi } }}\exp \left( { - \frac{{{A^2}}}{2}} \right) + \frac{1}{{\sqrt {2n\pi } \omega }}\exp \left( { - \frac{{{B^2}}}{2}} \right)
\end{equation}
where
\begin{align}\notag
A &= \frac{{{\gamma _{{\rm{ULAD}}}}}}{{\sqrt n }}\\\notag
B &= \frac{{{\gamma _{{\rm{ULAD}}}} - n - n \times E\left[ {\ln {z_i}|{H_1}} \right]}}{{\sqrt {n \times \tilde D\left[ {\ln {z_i}|{H_1}} \right]} }}\\\notag
\omega  &= \sqrt {\tilde D\left[ {\ln {z_i}|{H_1}} \right]}.
\end{align}
Let ${\frac{{\partial {P_{{\rm{error}}}}}}{{\partial {\gamma _{{\rm{ULAD}}}}}} = 0}$, we have
\begin{equation}
\alpha {\gamma _{{\rm{ULAD}}}}^2 + \beta {\gamma _{{\rm{ULAD}}}} + \mu  = 0
\end{equation}
where
\begin{align}\notag
\alpha  &= {\omega ^2} - 1\\\notag
\beta  &= 2n\left\{ {1 + E\left[ {\ln {z_i}|{H_1}} \right]} \right\}\\\notag
\mu  &= - {n^2}\left\{ {1 + {{\left\{ {E\left[ {\ln {z_i}|{H_1}} \right]} \right\}}^2} + 2E\left[ {\ln {z_i}|{H_1}} \right] + \frac{{{\omega ^2}}}{n}\ln {\omega ^2}} \right\}.
\end{align}

Notice that ${\Delta  = {\beta ^2} - 4\alpha \mu  > 0}$ holds at $\alpha\ne 0$ in practical CR system.  It can be proofed by means of  reduction to absurdity. By assuming  ${\Delta  \le  0}$, it can be obtained from (32) that ${\frac{{\partial {P_{{\rm{error}}}}}}{{\partial {\gamma _{{\rm{ULAD}}}}}} \le 0}$ (or ${\frac{{\partial {P_{{\rm{error}}}}}}{{\partial {\gamma _{{\rm{ULAD}}}}}} \ge 0}$) when ${\alpha  <0}$ (or ${\alpha  >0}$). In this case, the total error rate always decreases (or increases) with the detection threshold. Apparently, such cases discussed above are impossible in practical CR system, and it can be observed from the corresponding simulation results that the total error rate is not the monotonic function of the detection threshold. Therefore, we have ${\Delta  \ge 0}$ when $\alpha\ne 0$.

To obtain the optimal detection threshold $\gamma _{{\rm{ULAD}}}^*$, we give the following Proposition.

{\textbf{Proposition 3:}} On the basis of the quadratic equation in (32), the value of $\gamma _{{\rm{ULAD}}}$ corresponding to the minimum value of ${{P_{\rm{error}}}}$ can be given by
\begin{align}
\gamma _{{\rm{ULAD}}}^{\min } = \left\{ {\begin{array}{*{20}{c}}
{\frac{{ - \beta  + \sqrt \Delta  }}{{2\alpha }},\;\;\alpha  \ne 0}\\
{ - \frac{\mu }{\beta },\;\;\;\;\;\;\;\;\;\;\;\alpha  = 0}.
\end{array}} \right.
\end{align}

\emph{Proof.} The proof of Proposition 3 is detailed in Appendix-C.

Subject to the constraint on $P_f$, the closed-form expression for the optimal detection threshold can be obtained, given by
\begin{equation}
\gamma _{{\rm{ULAD}}}^* = \left\{ {\begin{array}{*{20}{c}}
{\frac{{ - \beta  + \sqrt \Delta  }}{{2\alpha }},\;\;{\rm{if}}\;\alpha  \ne 0,{P_f} = Q\left( {\frac{{ - \beta  + \sqrt \Delta  }}{{2\alpha \sqrt n }}} \right) \le {\zeta _{{P_f}}}}\\
{ - \frac{\mu }{\beta }{\rm{,\;\;\;\;\;\;\;\;\;\;\;\;\;if}}\;\alpha  = 0,{P_f} = Q\left( {\frac{{ - \mu }}{{\beta \sqrt n }}} \right) \le {\zeta _{{P_f}}}}\\
{{Q^{ - 1}}\left( {{P_f}} \right)\sqrt n ,{P_f} = 0.1\;,\;\;\;\;\;\;\;\;\;\;\;\;{\rm{otherwise}}.}
\end{array}} \right.
\end{equation}
Based on (34), the extra complexity of calculating ${\gamma _{{\rm{ULAD}}}^ * }$  can be ignored.

\section{Simulation results}\label{sec5}
In this section, we provide numerical results to validate the analyses on the detection performance and the optimal detection threshold for  the proposed ULAD. The results also illustrate the performance comparison of several detectors in the presence of the Laplacian noise. Without loss of generality, we assume ${\sigma _w^2 =1}$.

\begin{figure}[!t]
  \centering
    \includegraphics[width=0.4\textwidth]{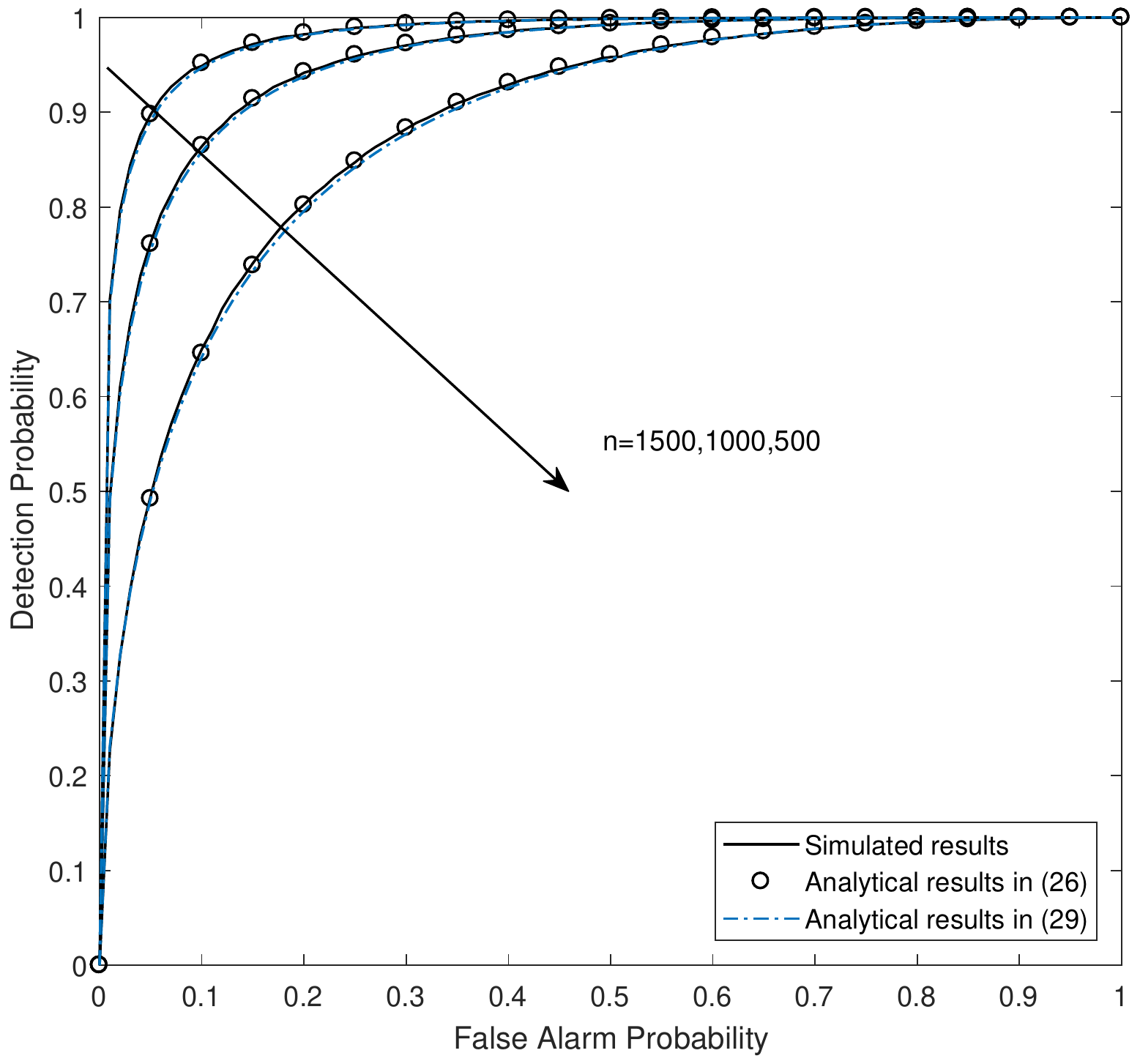}
  \caption{ROC curves of ULAD detector with different $n$ at ${\rho  =  - 14}$dB}
  \label{Fig. 2}
  \end{figure}

The Receiver Operating Characteristic (ROC) curves of ULAD are given in Fig. 2 at ${\rho  =  - 14}$dB for different sampling numbers (e.g., ${n}$=1500,1000,500). The analytical results are calculated by (26) and (29), where the infinite series factor $k$ is set as 1000 in (26). Obviously, the concurrence between simulated results and analytical results validates the theoretical analyses on detection and false alarm probabilities. Also, there is negligible approximation error in (29) compared with (26), and the detection probability converges to 1 faster with the rise of $n$.

The performance comparison of ULAD, POM, AVC, KS, AD, CM and ED in the presence of Laplacian noise are in Fig. 3. The $P_f$ is set to 0.05 and ${n}$=1000. As described in POM detector, its parameter $p$ is set as 0.05, 0.2 and 1.5. It can be observed that the proposed ULAD can achieve satisfactory performance under Laplacian noise as it takes full advantage of the statistical features of observations. For example, when ${\rho =-15}$dB, the proposed ULAD outperforms the POM ($p=$0.05), POM ($p=$0.2), AVC, KS, AD, CM, POM ($p=$1.5) and ED about 0.5dB, 1dB, 3.5dB, 3.8dB, 4.2dB, 4.2dB, 5dB and 6.5dB, respectively. In addition, the detection probability of the proposed detector is over 0.9 when  $\rho=-13$dB, while the others are below 0.85.

\begin{figure}[!t]
  \centering
    \includegraphics[width=0.4\textwidth]{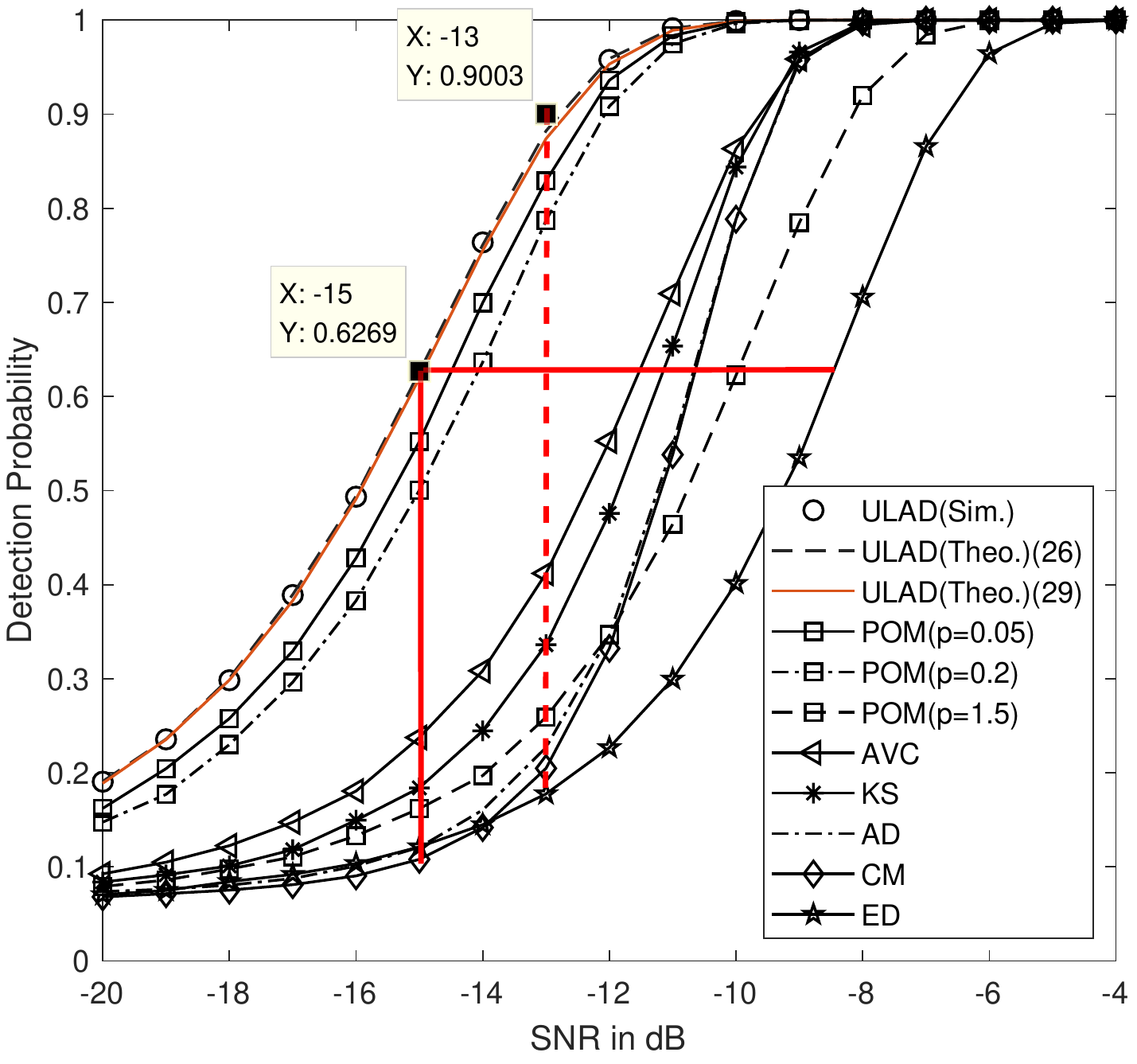}
  \caption{Detection probability versus ${\rho }$ of several sensing detectors under Laplacian noise with ${{P_f} = 0.05}$ at ${n=1000}$}
  \label{Fig. 3}
  \end{figure}

Fig. 4 compares the ROC curves of several detectors, i.e., ULAD, POM ($p=$ 0.05, 0.2, 1.5), AVC, KS, AD, CM and ED, at ${\rho  =  - 14}$dB with $n=$1000. It can be seen that the proposed ULAD shows superiority over other detectors, and the detection probability of ULAD goes to 1 much faster than others. For example, the detection probability of ULAD can achieve 0.9142 when $P_f=$0.15, whereas for other detectors $P_d$ are below 0.8 except POM ($p$=0.05,0.2).

\begin{figure}[!t]
 \centering
    \includegraphics[width=0.4\textwidth]{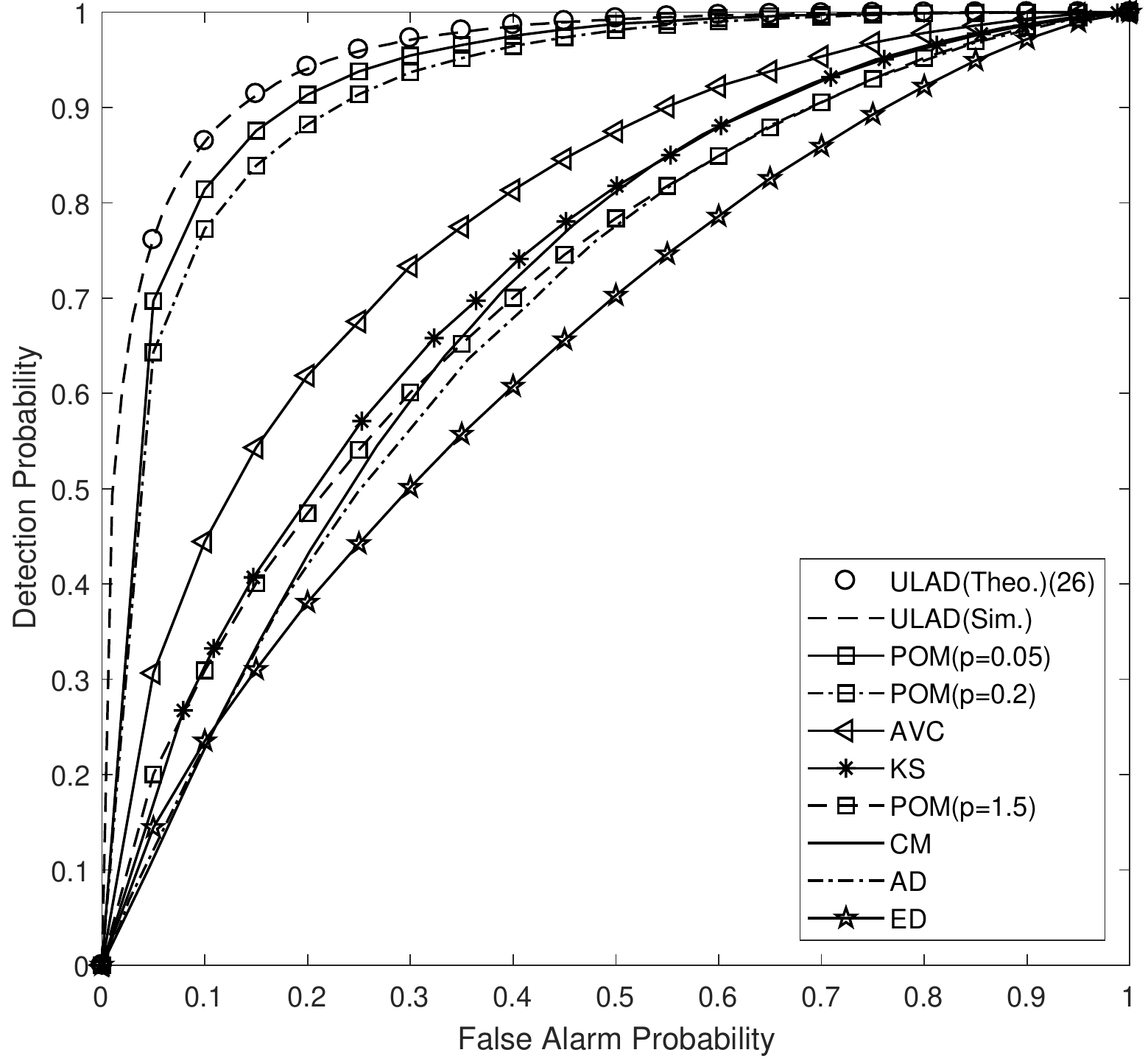}
  \caption{ROC curves of several sensing detectors at ${\rho  =  - 14}$dB with $n=1000$}
  \label{Fig. 4}
  \end{figure}

The performances of four detectors with two kinds of primary signals are shown in Fig. 5. The $P_f$ is set to 0.05, ${n}$=1000 and the parameter $p$ of POM detector is set as 0.05. It can be seen that comparing with fig. 3 where the BPSK primary signals is assumed, the performances of all detectors decrease when the primary signals is in sine waveform with single carrier frequency or assumed as a random uncorrelated Gaussian variable except the ED detector. Moreover, the proposed detector still outperforms others with different kinds of primary signals.

\begin{figure}[!t]
  \centering
    \includegraphics[width=0.43\textwidth]{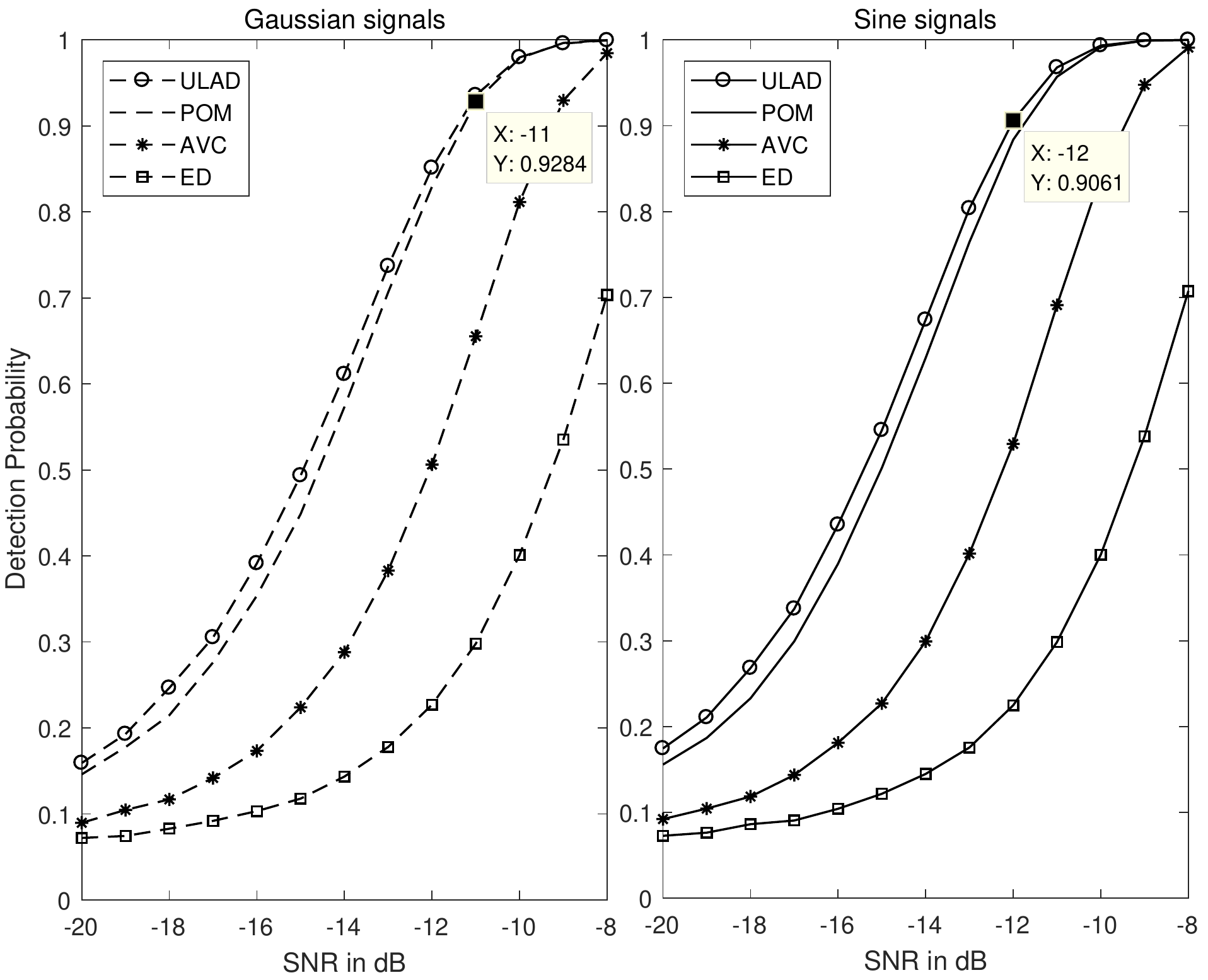}
  \caption{Detection probability versus ${\rho }$ of several detectors with different kinds of primary signals at ${{P_f} = 0.05}$ and ${n=1000}$}
  \label{Fig. 5}
  \end{figure}

\begin{figure}[!t]
 \centering
    \includegraphics[width=0.4\textwidth]{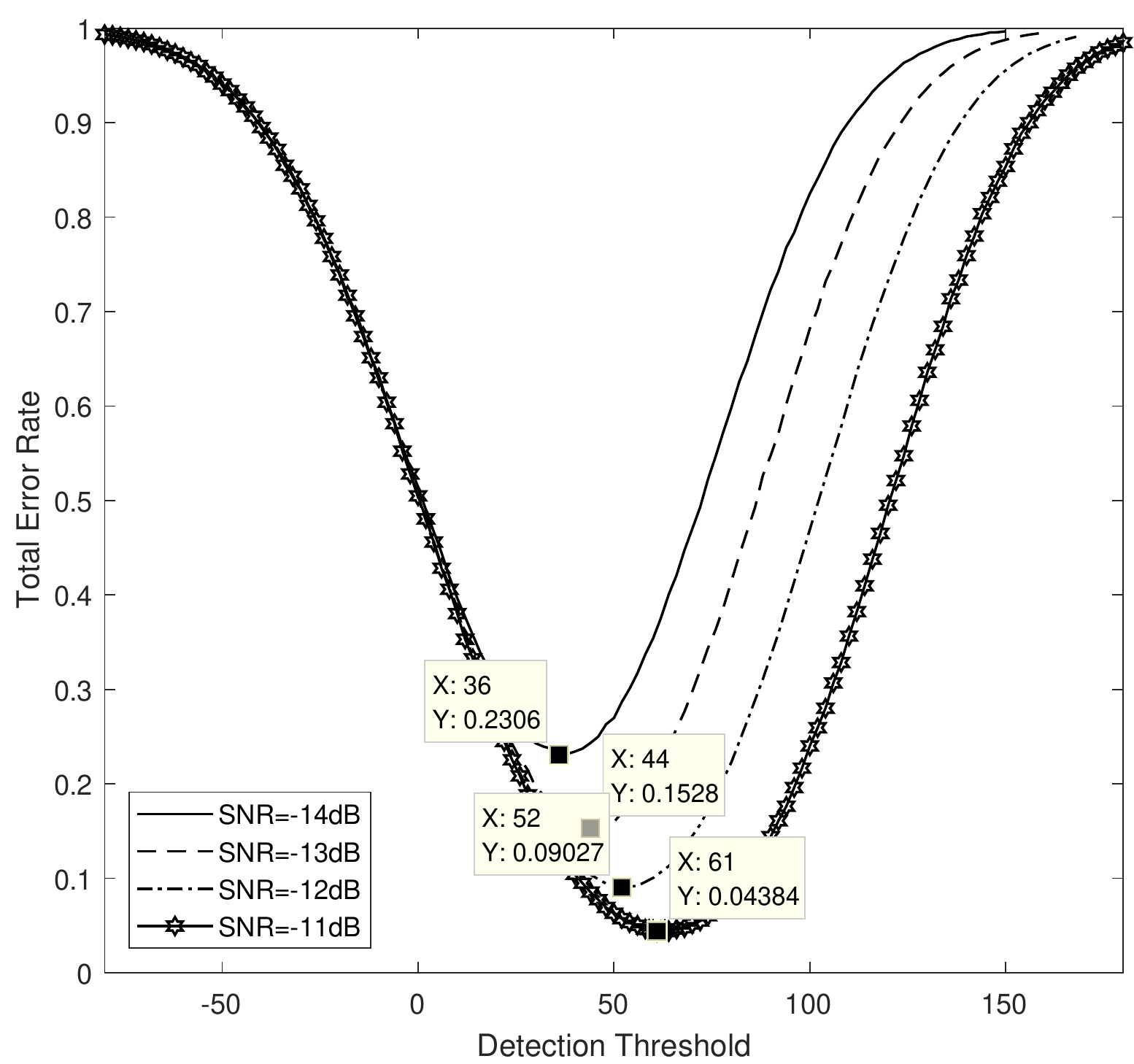}
  \caption{The total error rate against the detection threshold with different  ${\rho }$ at ${n=1000}$}
  \label{Fig. 6}
\end{figure}


\begin{table}[!t]
\centering
\caption{The optimal detection threshold for ${\xi _{{P_f}}}=0.1$ and $n=1000$}
\begin{tabular}{|c|c|c|c|c|}
\hline
 SNR & -14dB & -13dB & -12dB& -11dB \\
\hline
 $\gamma _{{\rm{ULAD}}}^ *$ & 40.5262 & 43.7242 & 51.9643 &61.4987 \\
\hline
$P_f$ & 0.1 & 0.0834 & 0.0502 & 0.0259 \\
\hline
\end{tabular}
\end{table}

Fig. 6 demonstrates the total error rate of the proposed ULAD against the detection threshold for different  ${\rho }$ at ${n=}$1000. The curves are obtained from ${ {10^5}}$ Monte Carlo simulations and the detection threshold varies from $-$80 to 180. Note that there exists the optimal detection threshold ${\gamma _{{\rm{ULAD}}}^ * }$ corresponding to the minimum total error rate at different ${\rho }$. For example, the optimal detection threshold ${\gamma _{{\rm{ULAD}}}^ * }$ are 36, 44, 52, 61 when ${\rho }$ are set to $-14$dB, $-13$dB, $-12$dB, $-11$dB, respectively. To verify the effectiveness of (34), the optimal detection threshold is calculated by (34), given in Table 2. It can be observed that the analytical thresholds coincide with the simulated thresholds well except one case that when ${\rho=-14}$dB. The reason is that the false alarm probability ${{P_f}}$ exceeds $0.1$ when ${\gamma _{{\rm{ULAD}}}^* =}$36. In such case, the constrain on (31b) can not be satisfied and the optimal detection threshold can be calculated by (22) at ${{P_f}=}$0.1.

Fig. 7 illustrates the comparison of the proposed ULAD versus ${\rho}$ with the optimal and fixed detection threshold (as ${{\gamma _{{\rm{ULAD}}}}}$ is decided by ${{P_f}}$ in (22), the fixed ${{P_f}}$ represents the fixed detection threshold.). The fixed ${{P_f}}$ is set to ${{10^{ - 1}}}$, ${{10^{ - 2}}}$, ${{10^{ - 3}}}$, ${{10^{ - 4}}}$, respectively and ${n=}$1000. It is shown that the fixed detection thresholds induce higher error as compared to the optimal detection threshold, and the detection performance of proposed ULAD can be improved considerably by means of the optimal threshold.
\begin{figure}[!t]
 \centering
    \includegraphics[width=0.38\textwidth]{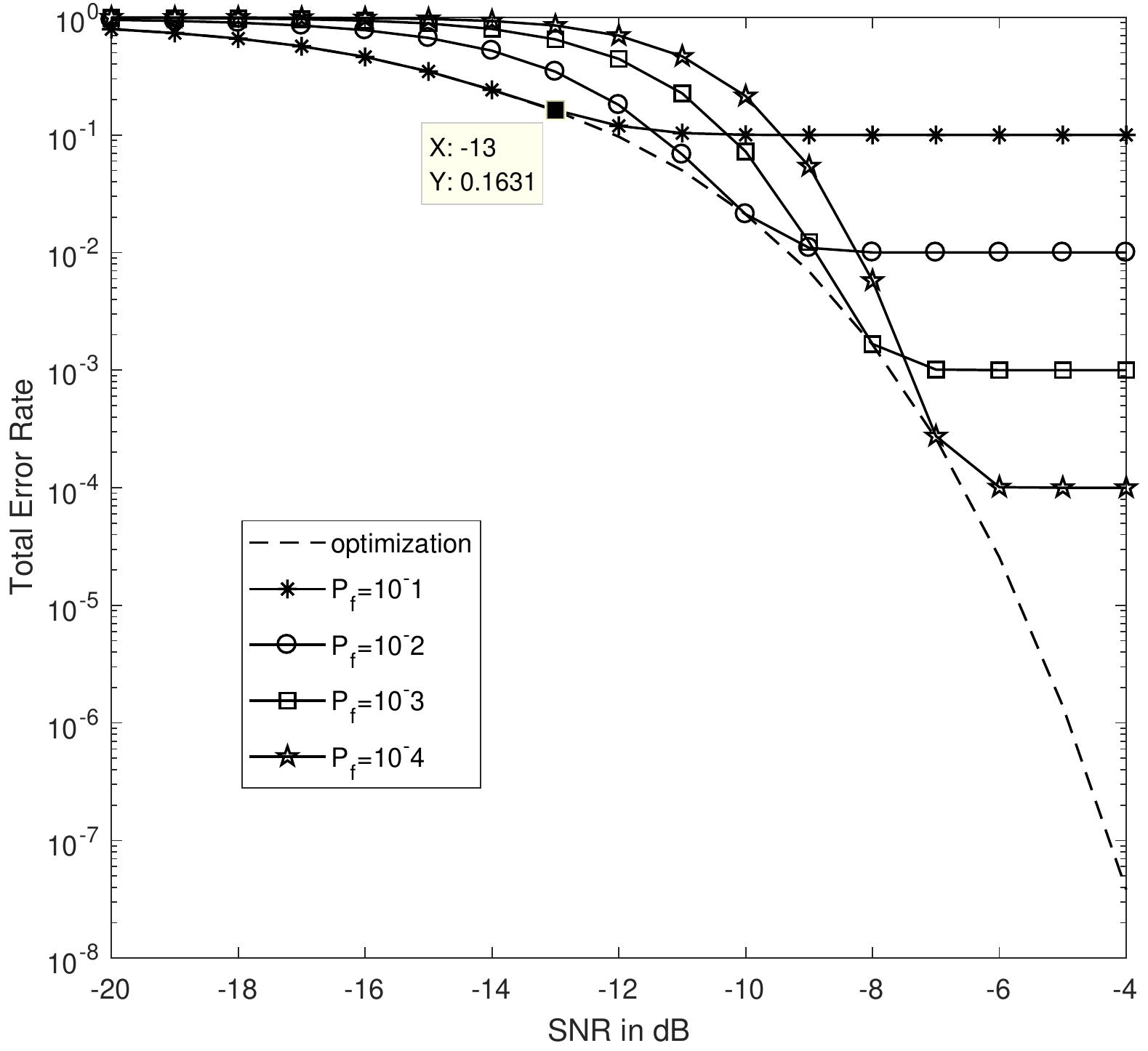}
  \caption{The total error rate versus different ${\rho}$ with the optimal and fixed detection threshold at ${n=1000}$}
  \label{Fig. 7}
  \end{figure}

\section{Conclusion}\label{sec6}
Exploiting the asymmetrical difference between the theoretical distribution and empirical distribution for the FLOM of received samples under Laplacian noise, we formulated the spectrum sensing problem as a unilateral GoF test problem, and then proposed a powerful and low complexity ULAD sensing detetctor.
The analytical expressions for the false alarm and detection probabilities were derived. On the basis of the derived expressions, the optimal detection threshold was  obtained.
It was shown that the proposed ULAD can achieve satisfactory detection performance in the presence of the heavy-tailed Laplacian noise, and outperform other existing detectors with the same considerations. In addition, the performance of ULAD was improved considerably with the optimal detection threshold as compared with the fixed detection threshold.

\section{Acknowledgements}\label{sec7}
The research was supported by  the National Key Research and Development Program of China (2016YFB1200202), the Natural Science Foundation of Shaanxi Province (2017JZ022), the Natural Science Foundation of China (61771365) and the 111 Project of China (B08038).

\section{References}\label{sec8}
\begin{enumerate}
\item[{[1]}] Federal Communications Commission., `Facilitating opportunities for flexible, efficient, and reliable spectrum use employing cognitive radio technologies, notice of proposed rule making and order', (FCC Document ET Docket, Dec, 2003), pp 03-322\vspace*{0.5pt}

\item[{[2]}] Haykin, S.: `Cognitive radio: brain-empowered wireless communications', {IEEE J. Sel. Areas Commun.}, 2005, 23, (2), pp 201-220\vspace*{0.5pt}

\item[{[3]}] Urkowitz, H.: `Energy detection of unknown deterministic signals', {Proc. IEEE}, 1967, 55, (4), pp 523-531\vspace*{0.5pt}

\item[{[4]}] Chen, Y.: `Improved energy detector for random signals in gaussian noise', {IEEE Trans. Wireless Commun.}, 2010, 9, (2), pp 558-563\vspace*{0.5pt}

\item[{[5]}] Cohen, D., Eldar, Y.C.: `Sub-Nyquist Cyclostationary Detection for Cognitive Radio', {IEEE Trans. Signal Process.}, 2017, 65, (11), pp 3004-3019\vspace*{0.5pt}

\item[{[6]}] Ayeh, E., Namuduri, K., Li, X.: `Performance evaluation of eigenvalue-based detection strategies in a sensor network'. {Proc. IEEE Int. Conf. Commun. (ICC)}, Sydney, 2014, pp 4407-4411\vspace*{0.5pt}

\item[{[7]}] Tan, F., Song, X., Leung, C., ~et al.: `Collaborative Spectrum Sensing in a Cognitive Radio System with Laplacian Noise', {IEEE Commun. Lett.}, 2012, 16, (10), pp 1691-1694\vspace*{0.5pt}

\item[{[8]}] Blackard, K.L., Rappaport, T.S., Bostian, C.W.: `Measurements and models of radio frequency impulsive noise for indoor wireless communications', {IEEE J. Sel. Areas Commun.}, 1993, 11, (7), pp 991-1001 \vspace*{0.5pt}

\item[{[9]}] Beaulieu, N.C., Niranjayan, S.: `{UWB} receiver designs based on a gaussian-laplacian noise-plus-{MAI} model', {IEEE Trans. Commun.}, 2010, 58, (3), pp 997-1006\vspace*{0.5pt}

\item[{[10]}] Marks, R.J., Wise, G.L., Haldeman, D.G., ~et al.: `Detection in Laplace Noise', {IEEE Trans. Aerosp. Electron. Syst.}, 1978, AES-14, (6), pp 866-872\vspace*{0.5pt}

\item[{[11]}] Li, Q., Li, Z., Shen, J., ~et al.: `A novel spectrum sensing method in cognitive radio based on suprathreshold stochastic resonance'. {Proc. IEEE Int. Conf. Commun. (ICC)} , Ottawa, June 2012, pp. 4426-4430\vspace*{0.5pt}

\item[{[12]}] Margoosian, A., Abouei, J., Plataniotis, K.N.: `An Accurate Kernelized Energy Detection in Gaussian and non-Gaussian/Impulsive Noises', {IEEE Trans. Signal Process.}, 2015, 63, (21), pp 5621-5636\vspace*{0.5pt}

\item[{[13]}] Zhu, X., Zhu, Y., Bao, Y., ~et al.: `A pth order moment based spectrum sensing for cognitive radio in the presence of independent or weakly correlated Laplace noise', {Signal Process.}, 2017, 137, pp 109-123\vspace*{0.5pt}

\item[{[14]}] Ye, Y., Li, Y., Lu, G., ~et al.: `Improved Energy Detection With Laplacian Noise in Cognitive Radio', {IEEE Syst. J.}, 2018, pp 1-12\vspace*{0.5pt}

\item[{[15]}] Gao, R., Li, Z., Li, H., ~et al.: `Absolute Value Cumulating Based Spectrum Sensing with Laplacian Noise in Cognitive Radio Networks', Wireless Pers. Commun., 2015, 83, (2), pp 1-18\vspace*{0.5pt}

\item[{[16]}] Ye, Y., Li, Y., Lu, G.: `Performance of Spectrum Sensing Based on Absolute Value Cumulation in Laplacian Noise'. {Proc. 2017 IEEE 86th Veh. Technol. Conf. (VTC-Fall)},  Toronto, Sept 2017, pp 1-5\vspace*{0.5pt}

\item[{[17]}] Wimalajeewa, T.,  Varshney, P.K.: `Polarity-Coincidence-Array Based Spectrum Sensing for Multiple Antenna Cognitive Radios in the Presence of Non-Gaussian Noise', {IEEE Trans. Wireless Commun.}, 2011, 10, (7), pp 2362-2371\vspace*{0.5pt}

\item[{[18]}] Karimzadeh, M., Rabiei, A.M., Olfat, A.: `Soft-Limited Polarity-Coincidence-Array Spectrum Sensing in the Presence of Non-Gaussian Noise', {IEEE Trans. Veh. Technol.}, 2017, 66, (2), pp 1418-1427\vspace*{0.5pt}

\item[{[19]}] Shao, M., Nikias, C.L.: `Signal processing with fractional lower order moments: stable processes and their applications', {Proc. IEEE}, 1993, 81, (7), pp 986-1010\vspace*{0.5pt}

\item[{[20]}] Re, E.D., Rupi, M.: `Comparison performance of efficient adaptive temporal filters and spatial arrays antennas based on fractional lower-order statistics in non-Gaussian environment'. {Proc. IEEE Veh. Technol. Conf. (VTC1999-FALL)}, Amsterdam, 1999, pp 1885-1889\vspace*{0.5pt}

\item[{[21]}] Wang, H., Yang, E.H., Zhao, Z., ~et al.: `Spectrum sensing in cognitive radio using goodness of fit testing', {IEEE Trans. Wireless Commun.}, 2009, 8, (11), pp 5427-5430\vspace*{0.5pt}

\item[{[22]}] Lei, S., Wang, H., Shen, L.: `Spectrum sensing based on goodness of fit tests'. {Proc. Int. Conf. Electron., Commun. Control (ICECC)}, Ningbo, Sept 2011, pp 485-489\vspace*{0.5pt}

\item[{[23]}] Ye, Y., Lu, G., Jin, M.: `Unilateral right-tail Anderson-Darling test based spectrum sensing for cognitive radio', {Electron. Lett.}, 2017, 53, (18), pp 1256-1258\vspace*{0.5pt}

\item[{[24]}] Rostami, S., Arshad, K., Moessner, K.: `Order-Statistic Based Spectrum Sensing for Cognitive Radio', { IEEE Commun. Lett.}, 2012, 16, (5), pp 592-595\vspace*{0.5pt}

\item[{[25]}] Jin, M., Guo, Q., Xi, J., ~et al.: `Spectrum sensing based on goodness of fit test with unilateral alternative hypothesis', {Electron. Lett.}, 2014, 50, (22), pp 1645-1646\vspace*{0.5pt}

\item[{[26]}] Li, Y., Ye, Y., Lu, G., ~et al.: `Cooperative spectrum sensing using discrete goodness of fit testing for multi-antenna cognitive radio system'. {Proc. Int. Conf. Comput., Inf. Telecommun. Syst. (CITS)}, Kunming, July 2016, pp 1-5\vspace*{0.5pt}

\item[{[27]}] Shen, L., Wang, H., Zhang, W., ~et al.: `Multiple Antennas Assisted Blind Spectrum Sensing in Cognitive Radio Channels', {IEEE Commun. Lett.}, 2012, 16, (1), pp 92-94\vspace*{0.5pt}

\item[{[28]}] Denkovski, D., Atanasovski, V., Gavrilovska, L.: `{HOS} Based Goodness-of-Fit Testing Signal Detection', {IEEE Commun. Lett.}, 2012, 16, (3), pp 310-313\vspace*{0.5pt}

\item[{[29]}] Shen, L., Wang, H., Zhang, W., ~et al.: `Blind Spectrum Sensing for Cognitive Radio Channels with Noise Uncertainty', {IEEE Trans. Wireless Commun.}, 2011, 10, (6), pp 1721-1724\vspace*{0.5pt}

\item[{[30]}] Zhang, G., Wang, X.,  Liang, Y.C., ~et al.: `Fast and Robust Spectrum Sensing via Kolmogorov-Smirnov Test', {IEEE Trans. Commun.} 2010, 58, (12), pp 3410-3416\vspace*{0.5pt}

\item[{[31]}] Gurugopinath, S., Muralishankar, R.,  Shankar, H.N.: `Differential Entropy-Driven Spectrum Sensing Under Generalized Gaussian Noise', {IEEE Commun. Lett.}, 2016, 20, (7), pp 1321-1324\vspace*{0.5pt}

\item[{[32]}] Sinclair, C.D., Spurr, B.D., Ahmad, M.I.: `Modified anderson darling test', { Commun. Stat. }, 2007, 19, (10), pp 3677-3686\vspace*{0.5pt}

\item[{[33]}] Nguyen-Thanh, N., Kieu-Xuan, T., Koo, I.: `Comments and Corrections Comments on "Spectrum Sensing in Cognitive Radio Using Goodness-of-Fit Testing"', {IEEE Trans. Wireless Commun.}, 2012, 11, (10), pp 3409-3411\vspace*{0.5pt}

\item[{[34]}] Tucker, H.G.: `A Generalization of the Glivenko-Cantelli Theorem', The Annals of Mathematical Statistics, 1959, 30, (3), pp. 828-830\vspace*{0.5pt}

\item[{[35]}] Gradshteyn, I.S., Ryzhik, I.M.: `Table of Integrals, Series, and Products (Seventh Edition)' (Academic press, Boston, 2007, 7th edn)\vspace*{0.5pt}

\item[{[36]}] He, Y., Ratnarajah, T.,  Yousif, E.H., ~et al.: `Performance analysis of multi-antenna GLRT-based spectrum sensing for cognitive radio', Signal Process., 2016, 120, pp 580-593\vspace*{0.5pt}

\item[{[37]}] {IEEE} 802.22 Wireless RAN: `Functional requirements for the 802.22 WRAN standard, {IEEE} 802.22-05/0007r46', Sept 2005
\end{enumerate}

\section{Appendix}\label{sec9}
\subsection{Appendix 1: Proof of Proposition 1}\label{subsec9.1}
From (14), we have ${z = {F_0}\left( x \right) = 1 - \exp \left( { - \sqrt {\frac{2}{{\sigma _w^2}}} x} \right)}$, $x \ge 0$, and its inverse function is given as
\begin{equation}
x =  - \sqrt {\frac{{\sigma _w^2}}{2}} \ln \left( {1 - z} \right),0 \le z \le 1.\tag{A.1}
\end{equation}
Then, the PDF of random variable ${z_i}$ under ${{H_0}}$ is given as
\begin{equation}
{g_0}\left( z \right) = \left| {\frac{{dx}}{{dz}}} \right| \cdot {f_0}\left( x \right) = \left\{ {\begin{array}{*{20}{c}}
{1,0 \le z \le 1}\\
{0,\;\;\;\;\rm{others}}.
\end{array}} \right.\tag{A.2}
\end{equation}
When ${H_0}$ is accepted, the mean and second-order origin moment of the variable ${\ln {z_i}}$ can be separately calculated as
\begin{align}\notag
E\left[ {\ln {z_i}|{H_0}} \right] &= \int_0^1 {\ln z \cdot {g_0}\left( z \right)} dz\\\notag
& = \left. {z\ln z} \right|_0^1 - \int_0^1 {zd\left( {\ln z} \right)} \\
& =  - 1\tag{A.3}
\end{align}
\begin{align}\notag
E\left[ {{{\left( {\ln {z_i}} \right)}^2}|{H_0}} \right] &= \int_0^1 {{{\left( {\ln z} \right)}^2} \cdot {g_0}\left( z \right)} dz\\\notag
 &= \left. {z \cdot {{\left( {\ln z} \right)}^2}} \right|_0^1 - 2\int_0^1 {\ln zdz}\\
& = 2.\tag{A.4}
\end{align}

The proof is complete.

\subsection{Appendix 2: Proof of Proposition 2}\label{subsec9.2}
Similar to (A.1) and (A.2), the PDF of ${z_i}$ under ${{H_1}}$ is given as
\begin{equation}
{g_1}\left( z \right) = \left| {\frac{{dx}}{{dz}}} \right| \cdot {f_1}\left( x \right) = \left\{ {\begin{array}{*{20}{c}}
{\frac{q}{{2{{\left( {1 - z} \right)}^2}}} + \frac{q}{2},\;0 \le z \le C}\\
{\frac{{{q^{ - 1}}}}{2} + \frac{q}{2},\;\;\;\;\;\;C < z \le 1}
\end{array}} \right.\tag{B.1}
\end{equation}
where ${q = \exp \left( { - \sqrt {\frac{{2\rho }}{{\sigma _w^2}}} } \right)}$ and ${C = 1 - q}$.

When ${H_0}$ is rejected, the mean of ${\ln {z_i}}$ can be given as
\begin{align}\notag
E\left[ {\ln {z_i}|{H_1}} \right] &= \int_0^1 {\ln z \cdot {g_1}\left( z \right)} dz\\\notag
&= \underbrace {\int_0^C {\ln z \cdot \frac{q}{{2{{\left( {1 - z} \right)}^2}}}} dz}_{\left( a \right)}\\
&+ \underbrace {\int_C^1 {\ln z \cdot } \frac{{{q^{ - 1}}}}{2}dz + \int_0^1 {\ln z \cdot \frac{q}{2}} dz}_{\left( b \right)}.\tag{B.2}
\end{align}
The first term of ($a$) can be calculated by
\begin{align}\notag
\left( a \right) &= \frac{q}{2}\int_0^C {\ln z \cdot \frac{1}{{{{\left( {1 - z} \right)}^2}}}} dz\\\notag
&=\left. {\frac{q}{2}\left[ {\frac{{\ln z}}{{1 - z}} - \ln {{\left( {\frac{{1 - z}}{z}} \right)}^{ - 1}}} \right]} \right|_0^C\\
&= \frac{q}{2}\left[ {\frac{{\ln C}}{{1 - C}} - \ln \left( {\frac{C}{{1 - C}}} \right)} \right]\tag{B.3}
\end{align}
and the second term of (${b}$) is given by
\begin{align}\notag
\left( b \right) &= \frac{1}{{2q}}\int_C^1 {\ln zdz}  + \frac{q}{2}\int_0^1 {\ln z} dz\\\notag
&= \frac{1}{{2q}}\left[ {z\ln z|_C^1 - \int_C^1 {zd\left( {\ln z} \right)} } \right] - \frac{q}{2}\\
 &= \frac{1}{{2q}}\left[ {C - C\ln C - 1} \right] - \frac{q}{2}.\tag{B.4}
\end{align}
By substituting (B.3) and (B.4) into (B.2), the ${E\left[ {\ln {z_i}|{H_1}} \right]}$ can be rewritten as
\begin{align}\notag
E\left[ {\ln {z_i}|{H_1}} \right] &= \frac{q}{2}\left[ {\frac{{\ln C}}{{1 - C}} - \ln \left( {\frac{C}{{1 - C}}} \right)} \right] \\
&+ \frac{1}{{2q}}\left[ {C - C\ln C - 1} \right] - \frac{q}{2}.\tag{B.5}
\end{align}
On the other hand, the second-order origin moment of the variable ${\ln {z_i}}$ under $H_1$ can be calculated by
\begin{align}\notag
E\left[ {{{\left( {\ln {z_i}} \right)}^2}|{H_1}} \right] &= \int_0^1 {{{\left( {\ln z} \right)}^2} \cdot {g_1}\left( z \right)} dz\\\notag
 &= \underbrace {\int_0^C {{{\left( {\ln z} \right)}^2} \cdot \frac{q}{{2{{\left( {1 - z} \right)}^2}}}dz} }_{\left( d \right)} \\\notag
 &+ \underbrace {\int_C^1 {{{\left( {\ln z} \right)}^2} \cdot } \frac{{{q^{ - 1}}}}{2}dz}_{\left( e \right)}\\
 &+ \underbrace {\int_0^1 {{{\left( {\ln z} \right)}^2} \cdot \frac{q}{2}} dz}_{\left( f \right)}.\tag{B.6}
\end{align}
The first term of (${d}$) is given by
\begin{align}\notag
\left( d \right) &= \frac{q}{2}\int_0^C {{{\left( {\ln z} \right)}^2} \cdot \frac{1}{{{{\left( {1 - z} \right)}^2}}}dz}\\\notag
&= \frac{q}{2}\int_0^C {{{\left( {\ln z} \right)}^2}d\left( {\frac{1}{{\left( {1 - z} \right)}}} \right)}  \\\notag
 &= \frac{q}{2}\left[ {\left. {\frac{{{{\left( {\ln z} \right)}^2}}}{{\left( {1 - z} \right)}}} \right|_0^C - 2\int_0^C {\ln z \cdot \left( {\frac{1}{{z\left( {1 - z} \right)}}} \right)dz} } \right]\\
 &= \frac{q}{2}\left[ {\underbrace {\left. {\frac{{{{\left( {\ln z} \right)}^2}}}{{\left( {1 - z} \right)}}} \right|_0^C - 2\int_0^C {\frac{{\ln z}}{z}dz} }_{\rm I} - \underbrace {2\int_0^C {\frac{{\ln z}}{{\left( {1 - z} \right)}}dz} }_{{\rm I}{\rm I}}} \right].\tag{B.7}
\end{align}
According to the Eq.(2.721.2) in [35], the first term of (${{\rm I}}$) can be simplified as
\begin{align}\notag
\left( {\rm I} \right) &= \left. {\frac{{{{\left( {\ln z} \right)}^2}}}{{\left( {1 - z} \right)}}} \right|_0^C - 2\left[ {\left. {\frac{{{{\left( {\ln z} \right)}^2}}}{2}} \right|_0^C} \right]\\
&= \frac{{{{\left( {\ln C} \right)}^2}}}{{\left( {1 - C} \right)}} - {\left( {\ln C} \right)^2}.\tag{B.8}
\end{align}
With the help of Eq.(2.727.2) and Eq.(2.728.2) in [35], the second term of (${{{\rm I}{\rm I}}}$) can be rewritten as
\begin{align}\notag
\left( {{\rm I}{\rm I}} \right) &= 2\left. {\left\{ { - \ln z\ln \left( {1 - z} \right)} \right\}} \right|_0^C + 2\int_0^C {\frac{{\ln \left( {1 - z} \right)}}{z}} dz\\\notag
 &=  - 2\ln C\ln \left( {1 - C} \right) - 2\left. {\left\{ {z \cdot \Phi \left( {z,2,1} \right)} \right\}} \right|_0^C\\
& =  - 2\ln C\ln \left( {1 - C} \right) - 2\mathop {\lim }\limits_{k \to \infty } \sum\limits_{i = 1}^k {\frac{{{C^i}}}{{{i^2}}}}\tag{B.9}
\end{align}
where ${\Phi \left( {z,s,v} \right) = \mathop {\lim }\limits_{k \to \infty } \sum\limits_{i = 0}^k {{{\left( {v + i} \right)}^{ - s}}{z^i}}}$,${\left[ {\left| z \right| < 1,v \ne 0, - 1, \ldots } \right]}$ is the Lerch function which is defined by Eq.(9.550) in [35].
By substituting (B.8) and (B.9) into (B.7), the ($d$) can be rewritten as
\begin{align}\notag
\left( d \right) & = \frac{q}{2}\left[ {\frac{{{{\left( {\ln C} \right)}^2}}}{{\left( {1 - C} \right)}} - {{\left( {\ln C} \right)}^2} + 2\ln C\ln \left( {1 - C} \right)} \right.\\\notag
&\left. { + \mathop {\lim }\limits_{k \to \infty } \sum\limits_{i = 1}^k {\frac{{2{C^i}}}{{{i^2}}}} } \right]\\\notag
 &= \frac{{qC}}{{2\left( {1 - C} \right)}}{\left( {\ln C} \right)^2} + q\ln C\ln \left( {1 - C} \right) + q\mathop {\lim }\limits_{k \to \infty } \sum\limits_{i = 1}^k {\frac{{{C^i}}}{{{i^2}}}}.\tag{B.10}
\end{align}
The second term of ($e$) can be given as
\begin{align}\notag
\left( e \right) &= \frac{1}{{2q}}\left[ {\left. {z{{\left( {\ln z} \right)}^2}} \right|_C^1 - 2\int_C^1 {\ln z} dz} \right]\\
 &= \frac{C}{q}\ln C - \frac{C}{{2q}}{\left( {\ln C} \right)^2} + \frac{{\left( {1 - C} \right)}}{q}\tag{B.11}
\end{align}
and the third term of ($f$) can  be  calculated by
\begin{equation}
\left( f \right) = \frac{q}{2} \cdot \int_0^1 {{{\left( {\ln z} \right)}^2}} dz = q.\tag{B.12}
\end{equation}
Substituting (B.10), (B.11) and (B.12) into (B.6), the second-order origin moment of the variable ${\ln {z_i}}$ under ${H_1}$ can be rewritten as
\begin{align}\notag
E\left[ {{{\left( {\ln {z_i}} \right)}^2}|{H_1}} \right]& = \frac{{C\left( {q - 1} \right)}}{{2q}}{\left( {\ln C} \right)^2} + q\ln C\ln q\\
&+ \frac{C}{q}\ln C + q\mathop {\lim }\limits_{k \to \infty } \sum\limits_{i = 1}^k {\frac{{{C^i}}}{{{i^2}}}} + 1 + q.\tag{B.13}
\end{align}

The proof is complete.

\subsection{Appendix 3: Proof of Proposition 3}\label{subsec9.3}
\emph{\textbf{Case 1:}} If ${\alpha  \ne 0}$, the roots of the quadratic equation in (32) can be obtained by
\begin{equation}
\gamma _{{\rm{ULAD}}}^{{\rm{zero}}} = \frac{{ - \beta  \pm \sqrt \Delta  }}{{2\alpha }}\tag{C.1}
\end{equation}
where $\Delta  = {\beta ^2} - 4\alpha > 0$.

(i) If ${\alpha  > 0}$, the total error rate will increase with the detection threshold when ${\gamma _{{\rm{ULAD}}}} < \frac{{ - \beta  - \sqrt \Delta  }}{{2\alpha }}$ or ${\gamma _{{\rm{ULAD}}}} > \frac{{ - \beta  + \sqrt \Delta  }}{{2\alpha }}$, and decrease when $\frac{{ - \beta  - \sqrt \Delta  }}{{2\alpha }} \le {\gamma _{\rm{ULAD}}} \le \frac{{ - \beta  + \sqrt \Delta  }}{{2\alpha }}$. Hence, the total error rate can achieve the minimum value when $\gamma _{\rm{ULAD}}^{\rm{min}}= \frac{{ - \beta  + \sqrt \Delta  }}{{2\alpha }}$.

(ii) If $\alpha  < 0$, the total error rate will decrease with the detection threshold when  ${\gamma _{{\rm{ULAD}}}} < \frac{{ - \beta  + \sqrt \Delta  }}{{2\alpha }}$ or ${\gamma _{{\rm{ULAD}}}} > \frac{{ - \beta  - \sqrt \Delta  }}{{2\alpha }}$, and increase when $\frac{{ - \beta  + \sqrt \Delta  }}{{2\alpha }} \le {\gamma _{\rm{ULAD}}} \le \frac{{ - \beta  - \sqrt \Delta  }}{{2\alpha }}$. Thus, the total error rate can achieve the minimum value when $\gamma _{\rm{ULAD}}^{\rm{min}}= \frac{{ - \beta  + \sqrt \Delta  }}{{2\alpha }}$.

In summary, if ${\alpha  \ne 0}$, the total error rate is able to achieve the minimum value at $\gamma _{\rm{ULAD}}^{\rm{min}}= \frac{{ - \beta  + \sqrt \Delta  }}{{2\alpha }}$.

\emph{\textbf{Case 2:}} If ${\alpha  = 0}$, the quadratic equation in (32) will transform into the following linear equation, given by
\begin{align}
\beta {\gamma _{{\rm{ULAD}}}} + \mu  = 0.\tag{C.2}
\end{align}
Obviously, we have
\begin{align}
\gamma _{{\rm{ULAD}}}^{\min } = \left\{ {\begin{array}{*{20}{c}}
{ - \frac{\mu }{\beta },\;\;\;\;\;\;\;\;\;\;\;\;\;\;\;\;\;\;\;\beta  > 0}\\
{{\rm{Not \;\;applicable}},\;\beta  \le 0}
\end{array}} \right..\tag{C.3}
\end{align}
Next we will prove that $\beta>0$ always holds and $\gamma _{{\rm{ULAD}}}^{\min } =  - \frac{\mu }{\beta }$ in what follows.

Based on (23) and (33), the first order derivative of $\beta$ with respect to $q$ can be written as
\begin{align}
\frac{{\partial \beta }}{{\partial q}} = 2n\frac{{\partial E\left[ {\ln {z_i}|{H_1}} \right]}}{{\partial q}} \tag{C.4}
\end{align}
where
\begin{equation}\notag
\frac{{\partial E\left[ {\ln {z_i}|{H_1}} \right]}}{{\partial q}} = \frac{{{q^2}\ln q + (1 - {q^2})\ln (1 - q) - {q^2} + q}}{{2{q^2}}}.
\end{equation}
Observe (C.4), we can see the sign of $\frac{{\partial \beta }}{{\partial q}}$ corresponds with the numerator term of $\frac{{\partial E\left[ {\ln {z_i}|{H_1}} \right]}}{{\partial q}}$. Therefore, we  just need to consider the sign of $f\left( q \right) = {q^2}\ln q + (1 - {q^2})\ln (1 - q) - {q^2} + q$.

The first derivative of $f\left( q \right)$ can be given as
\begin{equation}
\frac{{\partial f\left( q \right)}}{{\partial q}} = 2q\left( {\ln \frac{q}{{1 - q}} - 1} \right).\tag{C.5}
\end{equation}
It is apparent that the roots of $\frac{{\partial f\left( q \right)}}{{\partial q}} = 0$ are $q=0$ and $q=\frac{e}{{1 + e}}$. Meanwhile,
$f\left( q \right)$ will decrease with $q$ when $0 < q  < \frac{e}{{1 + e}} \approx   0.731$ and increases with $q$ for $ \frac{e}{{1 + e}}<q < 1$. Moreover, it can be obtained $f(0) = f(1) = 0$. Accordingly, the inequality $f(q) < \max \left\{ {f(0),f(1)} \right\}=0$ holds for $0 < q < 1$. Based on the above  analyses, we conclude that the inequality $\frac{{\partial E\left[ {\ln {z_i}|{H_1}} \right]}}{{\partial q}} <0$ holds for $0 < q < 1$, which indicates both $E\left[ {\ln {z_i}|{H_1}} \right]$ and $\beta$ decrease with the rise of $q$.
Due to ${E\left[ {\ln {z_i}|{H_1}} \right]}=-1$ at $q=1$, both ${E\left[ {\ln {z_i}|{H_1}} \right]}>-1$ and ${\beta  > 0}$ hold for $0 < q < 1$.

In conclusion, the value of $\gamma _{{\rm{ULAD}}}$ corresponding to the minimum value of ${{P_{\rm{error}}}}$ can be expressed as
\begin{align}
\gamma _{{\rm{ULAD}}}^{\min } = \left\{ {\begin{array}{*{20}{c}}
{\frac{{ - \beta  + \sqrt \Delta  }}{{2\alpha }},\;\alpha  \ne 0}\\
{ - \frac{\mu }{\beta },\;\;\;\;\;\;\;\alpha  = 0}
\end{array}} \right.\tag{C.6}
\end{align}

The proof is complete.

\end{document}